\documentclass[
 aps, pra,
 amsmath,amssymb,
 12pt,
 final,
tightenlines,
 twoside,
 onecolumn,
 nofloats,
nofootinbib,
 superscriptaddress,
showkeys,
showkeywords,
 ]
{revtex4-2}

\usepackage[T2A]{fontenc}
\usepackage[utf8]{inputenc}
\usepackage[russian,english]{babel}
\usepackage{graphicx}
\usepackage{dcolumn}
\usepackage{bm}

\usepackage{ulem}

\input{maik.rty}

\setcitestyle{authoryear,round}
\def\squareforqed{\hbox{\rlap{$\sqcap$}$\sqcup$}}

\def\sq{\ifmmode\squareforqed\else{\unskip\nobreak\hfil
\penalty50\hskip1em\null\nobreak\hfil\squareforqed
\parfillskip=0pt\finalhyphendemerits=0\endgraf}\fi}

\def\utw{\smash{\rlap{\lower5pt\hbox{$\sim$}}}}

\def\udtw{\smash{\rlap{\lower6pt\hbox{$\approx$}}}}

\def\diameter{{\ifmmode\mathchoice
{\ooalign{\hfil\hbox{$\displaystyle/$}\hfil\crcr
{\hbox{$\displaystyle\mathchar"20D$}}}}
{\ooalign{\hfil\hbox{$\textstyle/$}\hfil\crcr
{\hbox{$\textstyle\mathchar"20D$}}}}
{\ooalign{\hfil\hbox{$\scriptstyle/$}\hfil\crcr
{\hbox{$\scriptstyle\mathchar"20D$}}}}
{\ooalign{\hfil\hbox{$\scriptscriptstyle/$}\hfil\crcr
{\hbox{$\scriptscriptstyle\mathchar"20D$}}}}
\else{\ooalign{\hfil/\hfil\crcr\mathhexbox20D}}%
\fi}}

\def\be{\begin{equation}}
\def\ee{\end{equation}}
\def\ba{\begin{eqnarray}}
\def\ea{\end{eqnarray}}

\def\12{{1\over 2}}

\def\msun{M_\odot}
\def\lsun{L_\odot}

\def\ltsima{$\; \buildrel < \over \sim \;$}
\def\simlt{\lower.5ex\hbox{\ltsima}}
\def\gtsima{$\; \buildrel > \over \sim \;$}

\def\simgt{\lower.5ex\hbox{\gtsima}}

\usepackage[dvipsnames]{color}

\begin{document}
\selectlanguage{english}

\title{Interstellar Medium in Extremely High Star-Formation Regions: \\
A Prospect of Observations on the Millimetron Space Observatory\footnote{Published in Astronomy Reports, v. 69, pp. 913 -- 929 (2025), DOI~10.1134/S1063772925702154}}

\author{\firstname{E.~O.}~\surname{Vasiliev}}
 \email{eugstar@mail.ru}
 \affiliation{Lebedev Physical Institute, Russian Academy of Sciences, Moscow, 119991 Russia}

\author{\firstname{S.~A.}~\surname{Drozdov}}
 \affiliation{Lebedev Physical Institute, Russian Academy of Sciences, Moscow, 119991 Russia}

\author{\firstname{P.~V.}~\surname{Baklanov}}
 \affiliation{Lebedev Physical Institute, Russian Academy of Sciences, Moscow, 119991 Russia}
 \affiliation{National Research Center “Kurchatov Institute”, Moscow, 123182 Russia}

\author{\firstname{O.~P.}~\surname{Vorobyov}}
 \affiliation{Lebedev Physical Institute, Russian Academy of Sciences, Moscow, 119991 Russia}

\author{\firstname{S.~Yu.}~\surname{Dedikov}}
 \affiliation{Lebedev Physical Institute, Russian Academy of Sciences, Moscow, 119991 Russia}

\author{\firstname{M.~S.}~\surname{Kirsanova}}
 \affiliation{Lebedev Physical Institute, Russian Academy of Sciences, Moscow, 119991 Russia}
 \affiliation{Institute of Astronomy, Russian Academy of Sciences, Moscow, 119017 Russia}

\author{\firstname{T.~I.}~\surname{Larchenkova}}
 \affiliation{Lebedev Physical Institute, Russian Academy of Sciences, Moscow, 119991 Russia}

\author{\firstname{N.~N.}~\surname{Shakhvorostova}}
 \affiliation{Lebedev Physical Institute, Russian Academy of Sciences, Moscow, 119991 Russia}

\begin{abstract}
High star-formation rate and active galactic nucleus' emission can significantly transform the interstellar medium. In ultraluminous infrared galaxies, in which the star-formation rate reaches thousands of solar masses per year, the gas and dust are considerably affected by the ionizing radiation, cosmic rays and shock waves, that can be about a factor of 100--1000 larger than typical values in quiet star-forming galaxies. In these conditions, the emissivity of the gas and dust changes: in dense gas, high ionic and molecular transitions become excited, while dust grains are heated to high temperatures. In this paper, we analyze the possibilities for studying the interstellar medium in extreme conditions of ultraluminous infrared galaxies at redshifts of $\sim 0-3$, utilizing the atomic and molecular lines, and dust continuum in far infrared range of 100--500~$\mu$m.
We discuss the prospect of observations using the instruments of the Millimetron Space Observatory. 
\\
\\
\end{abstract}

\maketitle



\section{INTRODUCTION}
\label{vved} 

The processes of intense star formation in the central regions of galaxies produce large amounts of heavy elements (metals) and dust, which, mixing with gas, become part of the dense structure in the circumnuclear region -- the dust torus, that feeds the accretion disk \citep[e.g.,][]{Hickox2018}. The conditions required to sustain high rates of star formation and accretion in massive galaxies are still not fully understood.  The observations of such galaxies indicate a connection between the processes of accretion, the growth of supermassive black holes (SMBHs) \citep{Ho2004,Chen2013}, and the star formation rate \citep{Heckman2014}, which is expressed in the observed relation between SMBH mass and stellar population mass. The change of this relation is detected in galaxies at redshifts $z\simgt 6$ \citep{Decarli2018}. This raises numerous questions about the interrelation between the conditions in the interstellar medium and the efficiency of black hole growth.

Due to stellar activity, an accreting SMBH in a galactic center is fully or partially obscured by dust and gas. There is evidence that a significant fraction of growing black holes are obscured during their evolution \citep{Kelly2010}. Moreover, the proportion of dusty objects increases with decreasing luminosity \citep{Goulding2009}. Emission from an accreting SMBH passing through the torus is re-emitted in the infrared (IR) range. The ultraviolet (UV) radiation from stars is also efficiently absorbed and scattered by dust grains in the surrounding star-forming clouds. Thus, the IR radiation becomes the main source of information about both the accreting SMBH and star formation in the central regions of galaxies.

In this context, the conditions in the interstellar medium of massive luminous IR galaxies \citep{Sanders1996} remain puzzling. These are defined as objects with IR spectral luminosity ($8-1000\mu$m) exceeding the optical one \citep{Lagache2005}. Luminous infrared galaxies (LIRGs) and ultraluminous infrared galaxies (ULIRGs) have IR luminosities in the ranges of $10^{11} <L<10^{12}L_\odot$ and $10^{12}<L<10^{13}L_\odot$ \citep{Lagache2005,Casey2014}, respectively. In recent years, even more powerful galaxies, HyLIRGs (Hyper-LIRGs), with luminosities of $10^{13}<L<10^{14}L_\odot$ have been discovered \citep[e.g.,][]{Toba2016}. Very recently, an ELIRG (ExtremeLIRG) galaxy with IR luminosity $L>10^{14}L_\odot$ has been found \citep{Tsai2015}. The high luminosity of the whole LIRG class implies a very high star formation rate (SFR) -- orders of magnitude above that in our Galaxy -- and a high dust fraction, which obscures the galactic optical emission and reprocesses it into the IR range.

Luminous IR galaxies in the local Universe ($z<0.1$) are predominantly interacting (merging) systems and exhibit nuclear starbursts. During galaxy mergers, gas density, and thus the SFR, increases \citep{Kennicutt2012}, which is accompanied by rises in UV luminosity and the production rates of dust and metals. This leads to the obscuration of stellar clusters, reprocessing of stellar UV emission into IR photons, and the formation of luminous IR regions. The presence of black holes in colliding galaxies further boosts IR luminosity, as metal- and dust-enriched gas is readily accreted by the black hole.

Taking into acount that mergers and their induced starbursts span hundreds of millions of years, dust production can only be linked to Type II supernovae, since the other stellar sources operate on longer timescales. The direct observations of the historical supernovae indicate that Type II supernovae can eject up to $M_d\sim 1M_\odot$ of dust mass into the surrounding interstellar medium (ISM) \citep{Dunne2003,Temim2013,Indebetouw2014,Owen2015}. The balance between dust destruction by supernova shock waves \citep{McKee1989,Slavin2020,vs2024,Dedikov2025new,Dedikov2025azh} and replenishment of dust mass can determine the observed properties of luminous IR galaxies. Thus, the dust-to-stellar mass ratio in luminous IR galaxies can reflect the age of their starbursts.

It is worth noting that the average dust-to-stellar mass ratio in luminous IR galaxies is comparatively low ($\langle M_d/M_\ast\rangle \simeq 6\times 10^{-4}$ \citep{U2012}) and even may be about the order of the value in our Galaxy. However, variations in $M_d/M_\ast$ for individual galaxies may reach an order of magnitude, suggesting short dust lifetimes in luminous IR galaxies due to the intense energy release from supernovae. For comparison, in the ISM of our Galaxy the dust destruction timescale by supernova shocks is only about 100~Myr \citep[see discussion in][]{Draine2003}. Since the supernova rate in the Milky Way is 1.5 -- 2 orders of magnitude lower than in luminous IR galaxies, we can expect dust lifetimes proportionally shorter in the latter, i.e., about 1~Myr.

Activity of galactic nuclei is linked to the helium reionization in the Universe at redshifts 2 -- 3 \citep[see, e.g.][]{Ferrara2014}. The estimates of quasar luminosity functions \citep{Wolf2003,Masters2012,Palanque2016} compared to the star formation history models reveal a deficit of quasars \citep[see, e.g.][]{Ferrara2014}. It is natural to suppose that some quasars are obscured by dust. Dusty IR galaxies have been discovered in a blind sky survey with the SCUBA camera at redshifts $z\sim 2-4$ \citep{Barger1998} and are characterized by rather high star formation rate of $100-1000~\rm M_{\odot}$/yr \citep{Casey2014}, which is comparable to typical values for luminous IR galaxies in the local Universe.

The detection of SMBHs with masses $M_{BH}\sim 10^8-10^9~\msun$ at redshifts greater than 7 \citep{Mortlock2011,Banados2018,Wu2015,Yang2020}, and of substantial dust and metal masses in such early galaxies \citep{Decarli2018,Venemans2019,Decarli2022}, has only deepened the puzzles of rapid SMBH growth and star formation efficiency. This demands more refined numerical approaches to modeling the evolution of the medium around SMBHs \citep{Fan2019,Petric2019}.

It can be noted that galaxies with high IR luminosity of $\sim 10^{12-14}\lsun$ and star formation rates up to several thousand $\msun$/yr are observed across a wide redshift range: in the local Universe \citep{Sanders1996,Lonsdale2006} and during the epoch of helium reionization $z\sim 2-3$ \citep{Casey2014}. At $z\simlt 0.3$ such objects are rare -- fewer than one per nearly 100 square degrees. At $z\sim 0.3-1$ their number rises sharply, reaching densities of several hundred per square degree at $z\simgt 1$ \citep[e.g.,][]{RowanRobinson1997}. Recently, luminous IR galaxies with star formation rates of nearly $1500\msun$/yr have been observed before the onset of hydrogen reionization at $z\simgt 7$ \citep{Venemans2019}.

The similarity of galaxies with high IR luminosity across redshifts is reflected in the relation between their far-IR luminosities and the C II line \citep{Decarli2018}. Most of the IR emission of such galaxies appears to be powered by star formation \citep[e.g.,][and others]{Genzel1998}. This raises the question of what conditions in the interstellar medium sustain such extreme SFRs. These conditions are clearly associated with intense radiation and cosmic ray fluxes, modifications of ionization and molecular kinetics, and efficient dust heating and destruction. Since SMBH evolution and high SFRs are apparently connected, studying the properties of the ISM in galaxies with extreme star formation will provide better insight into SMBH growth processes.

In the near future, two major space telescopes will operate in the IR spectral range (inaccessible to ground-based instruments due to atmospheric absorption): the recently launched James Webb Space Telescope (JWST) \citep{Kalirai2018} and the planned Millimetron space observatory, scheduled for launch in the early 2030s \citep{Kardashev2014,Novikov2021,Likhachev2024}.

For the high-resolution spectrometer of the Millimetron observatory, seven bands M1 -- M7 are proposed: 500 -- 600, 740 -- 900, 1080 -- 1230, 1300 -- 1400, 1890 -- 1910, 2390 -- 2410, and 2660 -- 2680 GHz \citep{Tretyakov2025}. The sensitivity estimates of the high-resolution spectrometer are determined by the noise temperature of the heterodyne receivers \citep{Tretyakov2025}. For example, for band M1 in the range 500 -- 600 GHz (500 -- 600~$\mu$m) $T_{sys} \simeq 200$K, which allows a sensitivity of $\sim 2\times 10^{-18}$ W~m$^{-2}$ or $\sim 3$~mJy and a ratio $S/N=5$ at a spectral resolution $R=10^5$ in one hour of observations\footnote{https://millimetron.ru/dlya-uchenykh/kalkulyator-chuvstvitelnosti}. For the 10-m mirror of the Millimetron observatory, the angular resolution will be 15'' at a wavelength of 500~$\mu$m.

The capabilities of Millimetron are especially promising for studying the interstellar medium in luminous IR galaxies both in the local Universe and at high redshifts, since its working wavelength range (100-500~$\mu$m) covers the main IR lines of molecules, ions, and metal atoms, along with a significant portion of the dust continuum. We will consider these capabilities here.  Section 2 briefly describes the properties of nearby ultraluminous IR galaxies Arp 220 and UGC 5101. Section 3 analyzes prospects for observing IR lines of ions and metal atoms. Section 4 addresses molecular lines. Section 5 discusses the dust continuum and molecular absorption lines. The conclusion outlines tasks for the spectral instruments of the Millimetron space observatory.

\section{GALAXIES Arp 220 AND UGC 5101}

Dozens of ultraluminous infrared galaxies (ULIRGs) have been discovered in the nearby Universe ($z\simlt 0.3$) \citep{Lonsdale2006}. A significant fraction of these objects show evidence of major galactic mergers, active nuclei, and regions of intense star formation. At the same time, the brightest regions of infrared emission turn out to be fairly compact and Compton-thick, $N_{\rm H} \simgt 10^{24}$cm$^{-2}$ \citep{Condon1991,Hickox2018}.

The best-known and closest representative of ULIRGs in the local Universe is Arp 220 \citep{Soifer1984,Joy1986}, located at a distance of 77~Mpc ($z\simeq 0.018$), apparently formed as a result of a merger, and containing two AGNs \citep{Norris1988,Graham1990,Clements2002}. The two nuclei—Arp 220 East and Arp 220 West -- are located in the center of the galaxy at a distance of about $\sim 380$ pc from each other. Each nucleus is surrounded by its own dusty gas disk, while at the same time there exists a common kiloparsec-scale dusty molecular gas disk. The mass of each black hole is estimated at $\sim 10^5\msun$.

The total IR luminosity is estimated at $L_{IR} \sim 2 \times 10^{12} L_{\odot}$ \citep{Rangwala2011}. Most of the far-IR and submillimeter emission of Arp 220 is generated in a central region of size $R_{IR} \sim 300$ pc. Observations of CO emission and rotational lines of HCN and HCO$^+$ indicate that the central region is very inhomogeneous in density, consisting of a mixture of "diffuse" low-density gas and dense molecular condensations, or cloudlets. The radius of these cloudlets is estimated in the range $r_{cl} \sim 0.03 - 0.3$ pc.

Millimeter-wavelength observations reveal a large mass of dust within 50 pc of the active nuclei \citep{Scoville2015}, along with high volume and column density of molecular hydrogen: $\sim 10^5$cm$^{-3}$ и $\sim 10^{22}$cm$^{-2}$, respectively \citep{Scoville2015,Sakamoto2008}. Unsurprisingly, emission lines of highly excited CO and HCN molecules, as well as OH$^+$ and H$_2$O, have been detected under such conditions \citep{Rangwala2011}. These molecules form in dense gas within dissociation regions exposed to UV and X-ray radiation \citep{Wolfire2022}. Under such conditions, high-level transitions in water molecules are excited, the energy levels up to $800-1000$~K can be populated \citep{Liu2017}; the dust temperature can reach $\sim 100-200$~K. The analysis of the level populations in such an extreme interstellar medium shows that the relative abundance of water molecules varies from $10^{-9}-10^{-8}$ in cold regions to $10^{-8}-10^{-7}$ in warm ones, reaching $10^{-6}-10^{-5}$ in hot regions due to efficient release of molecules at grain surfaces into the gas phase at such high temperatures.

The dust-to-stellar mass ratio in the Arp 220 galaxy is $\zeta_{d}^\ast=M_d/M_\ast\simeq 0.002$ \citep{U2012}, which is four times higher than in our Galaxy. The dust-to-gas mass ratio is also slightly higher than in the Galaxy: $\zeta_d=M_d/M_{\rm H}\sim 0.01-0.02$ \citep{Scoville1997,Rangwala2011}. Therefore, it is most likely that the bulk of the dust in Arp 220 was produced during its starburst phase. To determine the sources of such a large amount of dust, estimates of the timescales of dynamical processes are required. First, the optical observations indicate the presence of two stellar populations: a young one with an age $t_{_{\rm YSP}}\leq 10$ Myr and an intermediate one with $t_{_{\rm ISP}}\sim 300$ Myr \citep{Wilson2006}. The young population is concentrated in the more compact central region of Arp 220, contributing up to over 60\% of the total luminosity \citep{Rodrigez2008}. Second, for the size of the active region of Arp 220 $R_{\rm IR}\sim 1$ kpc \citep{Emerson1984} and a characteristic merger velocity $v_m\geq 100$~km~s$^{-1}$ the dynamical timescale is $t_{dyn}\leq 10$ Myr. On such timescales, only Type II supernovae can be effective dust sources. In this case, the estimated dust yield per supernova is several times greater than that measured in SN 1987A $\simeq 0.2M_\odot$ \citep{Indebetouw2014}. One might assume that the duration of the starburst needs to be longer, especially since other estimates suggest the young and intermediate populations have ages $t_{_{\rm YSP}}\leq 100$ Myr and $0.5\leq t_{_{\rm ISP}}\leq 0.9$ Gyr, respectively \citep{Rodrigez2008}. However, this contradicts what follows from the merger velocity. Moreover, it is rather difficult to sustain such a high rate of star formation over tens of millions of years.

The galaxy UGC~5101, located at a distance of nearly 160~Mpc ($z\simeq 0.038$), belongs to the class of galaxies with strong low-ionization nuclear emission line regions (LINERs) \citep{Veilleux1995}. Dust absorption is fairly strong, although somewhat less than in Arp~220 \citep{Rigopoulou1996}. At the same time, a significantly greater number of metal ion lines are observed in UGC~5101 than in Arp~220 \citep{Armus2004}. The flux in the [CII] 158~$\mu$m line is comparable, despite nearly twice the distance and roughly half the total IR luminosity of Arp~220, $\sim 10^{12}\lsun$.

Despite numerous observations of Arp 220 and UGC 5101, fundamental questions remain about the nature of their powerful IR emission, its maintenance, the duration of their starbursts, and many other issues. Similar problems arise when studying other local and distant infrared galaxies. We will consider possible indicators of physical processes in the interstellar medium of such galaxies and the observational capabilities of the instruments of the Millimetron space observatory.

\begin{figure}
\center
\includegraphics[width=170mm]{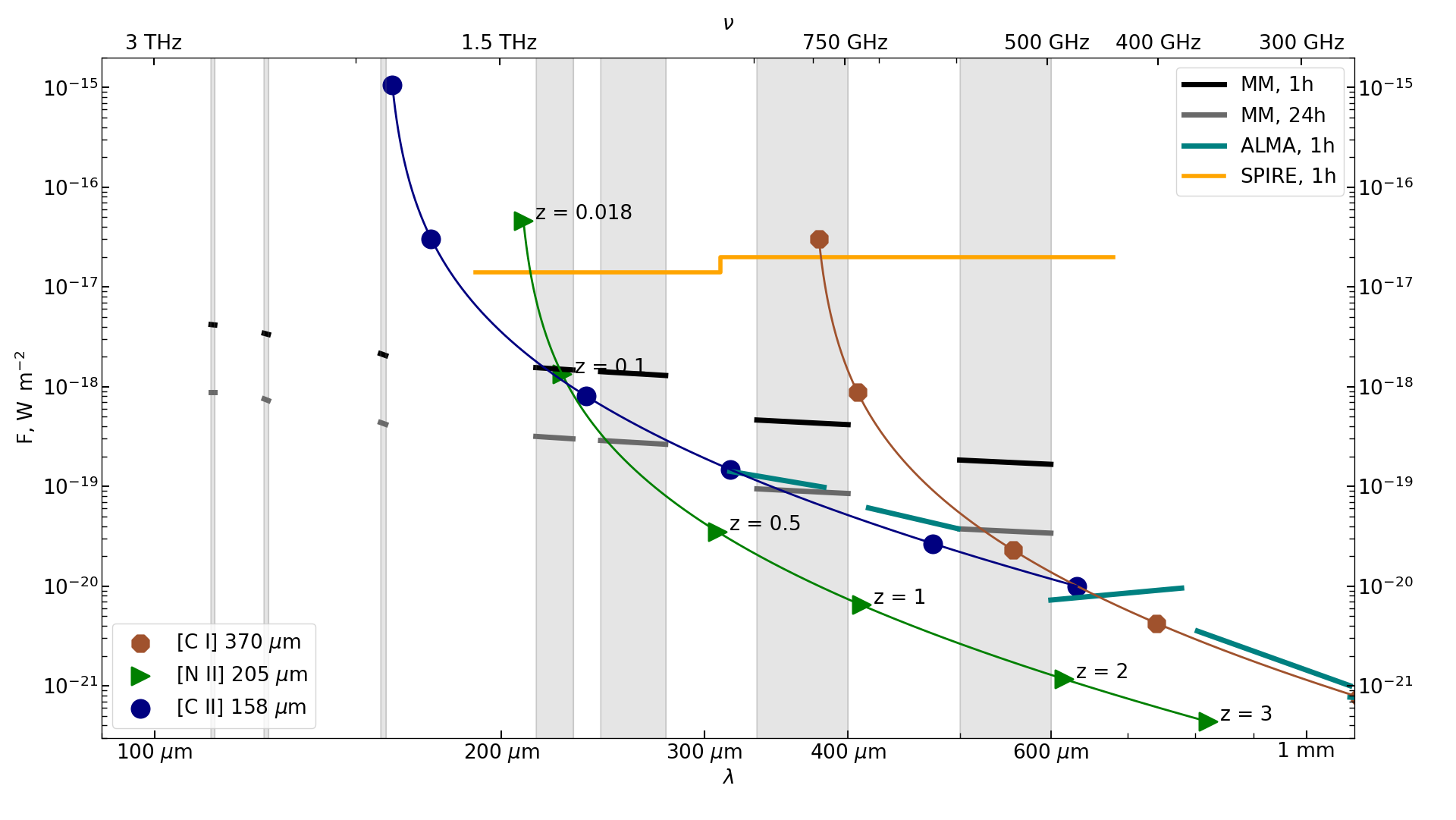}
\caption{ 
Expected fluxes in the lines of metal ions from a galaxy similar to Arp 220 at redshift $z$. The symbols from top to bottom correspond to the redshift of the galaxy Arp~220 $z=0.018$ and further $z=0.1$, 0.5, 1, 2 and 3 (for reference these values are indicated near the symbols along one of the curves). The fluxes for $z=0.018$ correspond to those observed in the galaxy Arp 220 and are scaled to the total luminosity in the IR range $10^{12}\lsun$ \citep{Rangwala2011}. Gray vertical bands show the wavelength ranges of operation of the high-resolution spectrometer of the Millimetron observatory. Thick black lines correspond to the $5\sigma$ receiver sensitivity for an integration time of 1~h and spectral resolution $R = 10^3$. Thick gray lines correspond to 24~h of integration. Thick green lines show the sensitivity for the ALMA interferometric array observations in the 50-antenna mode at $R=10^3$ and 1~h integration, with significance of $5\sigma$. The thick yellow line corresponds to the Herschel/SPIRE sensitivity at 1~h integration with resolution $R=10^3$.
}  
\label{fig-arp220-ions}
\end{figure}

\section{IONIZED GAS}

The spectral instruments of the Millimetron observatory \citep{Kirsanova2025,Tretyakov2025} can cover several diagnostic lines of metal ions from objects within redshift range $z\sim 0-5$: [CII] 158~$\mu$m, [OI] 63 и 145~$\mu$m, [NII] 122 и 205~$\mu$m, [OIII] 52 и 88~$\mu$m, [NIII] 57~$\mu$m, [CI] 369~$\mu$m and others. These transitions mainly arise in gas within photodissociation regions and HII ionized regions. The [NIII] 57~$\mu$m/[NII] 122~$\mu$m ratio characterizes the hardness of the UV radiation spectrum \citep{Allen2008}. Lines of [NIII] and [OIII] can be used to study the production of nitrogen and oxygen in intermediate-mass stars \citep{Spinoglio2022}. These lines are also used to determine the gas metallicity in regions with low dust absorption in ultraluminous galaxies \citep{Chartab2022}.

Spectral observations of the galaxy Arp 220 with the ISO orbital space telescope revealed a reduced [CII]/IR luminosity ratio compared to normal starforming galaxies. One explanation for the observed [CII] deficit is that the interstellar medium of Arp 220 is optically thick in the far-infrared \citep{Scoville1991,Fischer1997}.However, the estimates of the extinction are insufficient to explain the absence of [OIII] 88~$\mu$m emission in Arp 220 \citep{Fischer1997}. Due to the high absorption, the emission from the ionized gas in Arp 220 is limited compared to some other luminous IR galaxies. It is possible that the higher spatial resolution of the Millimetron space observatory will allow us to study the of ISM properties in the vicinity of the two SMBHs.

Galaxies like Arp 220 can be studied in the [NII] 122~$\mu$m and [CII] 158~$\mu$m lines up to $z\simlt 0.5$ (Fig.~\ref{fig-arp220-ions}). However, the most valuable information may come from observations at $z\simlt 0.1$, which provide insights into the neutral ISM of luminous IR galaxies. These lines characterize the transition from warm neutral to diffuse ionized ISM phases. In Arp 220, the transition regions are obscured by a dense dust layer, and the line flux ratios serve as an additional diagnostic of absorption and, consequently, dust properties.

\begin{figure}
\center
\includegraphics[width=170mm]{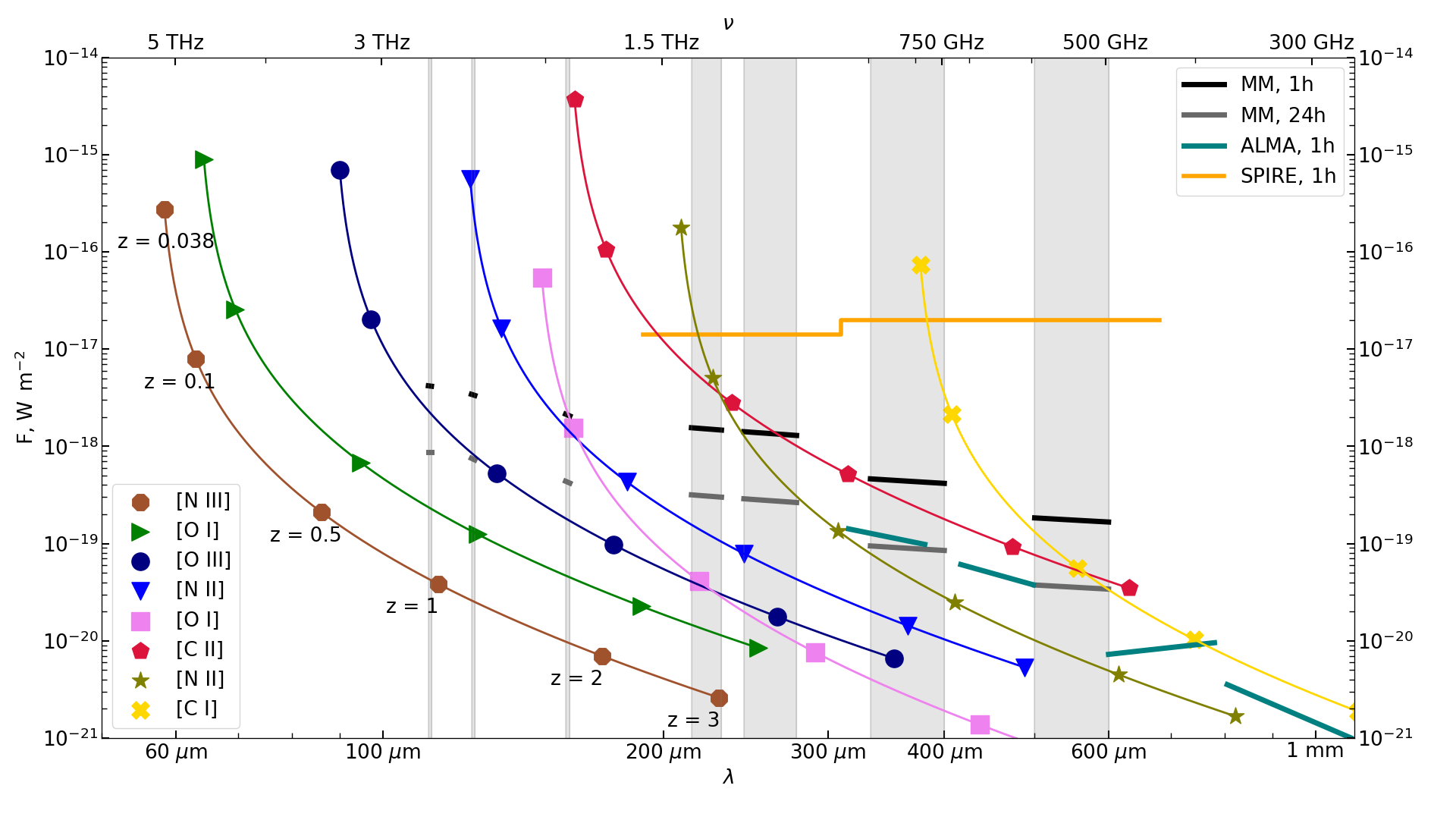}
\caption{ 
Expected fluxes in the lines of metal ions from a galaxy similar to UGC 5101 at redshift $z$. The symbols from top to bottom correspond to the redshift of the galaxy UGC 5101 $z=0.038$ and further $z=0.1$, 0.5, 1, 2 and 3 (for reference these values are indicated near the symbols along one of the curves). The fluxes for $z=0.038$ correspond to those observed in the galaxy UGC 5101 and are scaled to the total luminosity in the IR range $10^{12}\lsun$ \citep{Armus2004,Spinoglio2015}. Other notations are the same as in Fig.~\ref{fig-arp220-ions}.
}  
\label{fig-ugc5101-ions}
\end{figure}

Not all luminous IR galaxies exhibit such high extinction at wavelengths $\lambda \sim 100-500$~$\mu$m. In particular, the galaxy UGC 5101 shows significantly more metal-ion lines than Arp 220 (Fig.~\ref{fig-ugc5101-ions}). It should be noted that the [CII] 158~$\mu$m ine flux is comparable, despite UGC 5101 being nearly twice as distant and having roughly half the total IR luminosity $\sim 10^{12}\msun$. However, this does not imply much lower ISM absorption, since in the direction of the central black hole in UGC 5101 the gas is Compton-thick, with a column density exceeding $10^{24}$~cm$^{-2}$ \citep{Oda2017}. Here, X-ray radiation from the central SMBH of mass $\sim 10^8\msun$ plays a more significant role than in Arp 220, where star formation processes dominate. In this sense, luminous IR galaxies include starburst galaxies, Seyfert galaxies, and galaxies with strong low-ionization metal emission lines (LINERs), and the galaxy UGC 5101 belongs to the latter class. Thus, powerful IR radiation, dense star-forming regions, and high star formation rates can occur in galaxies of various types. Ratios of metal-ion lines can help identify the primary ionization source and clarify the origin of regions with extreme star formation.

Figure~\ref{fig-ugc5101-ions} shows the expected metal-ion line fluxes from a galaxy like UGC 5101 at different redshifts. It can be seen that a considerable number of lines can be detected up to $z\simlt 0.5$. The flux ratio of [OIII] 88~$\mu$m и [NII] 122~$\mu$m is a sensitive indicator of gas density and ionization parameter $U$ \citep{Spinoglio2022}, which can be used to estimate the production of ionizing photons by various radiation sources \citep{DiazSantos2017}.

\begin{figure}
\center
\includegraphics[width=170mm]{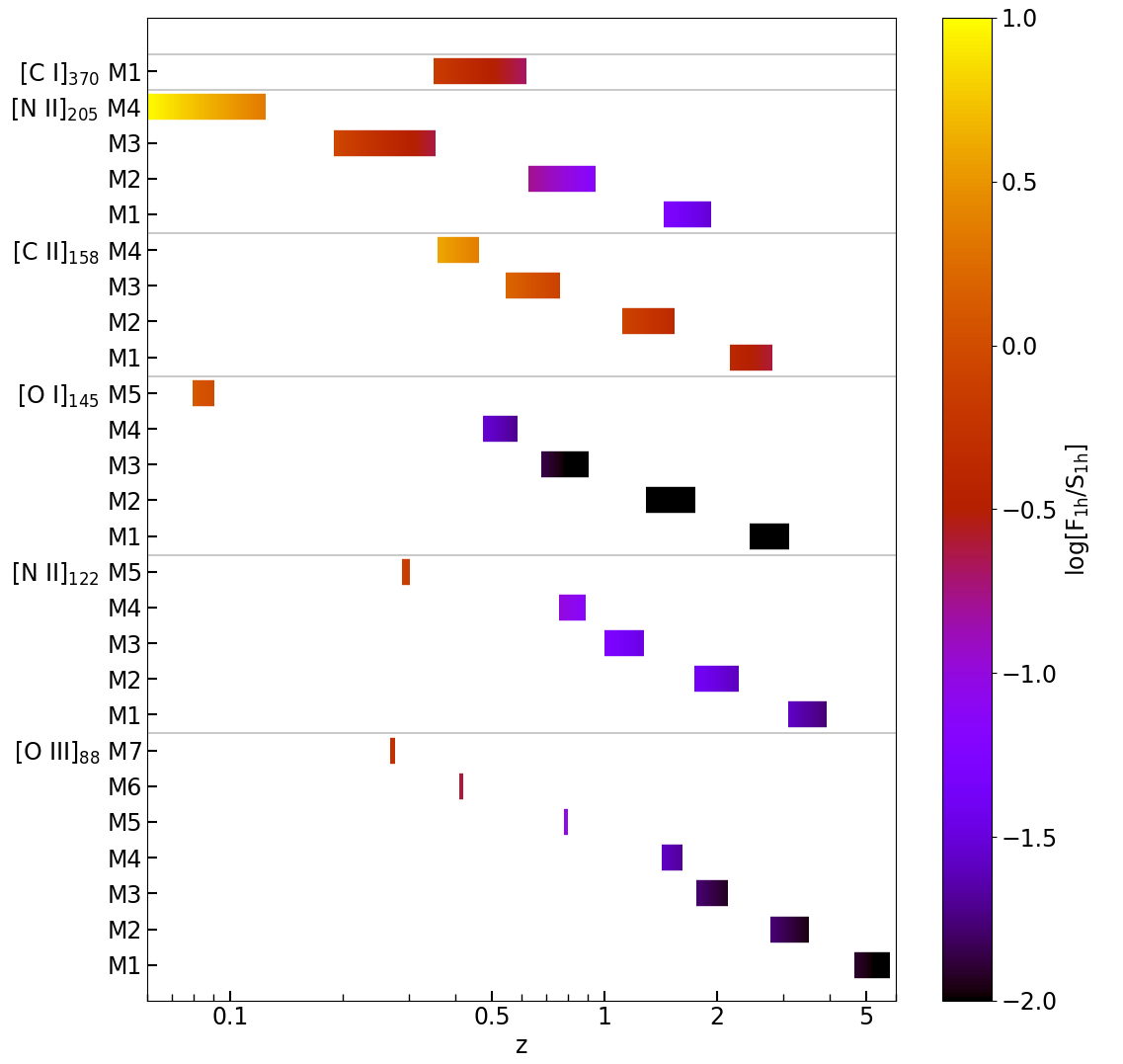}
\caption{ 
Observational capabilities for several lines of metal ions from an object at redshift $z$ in bands M1--M7 of the high-resolution spectrometer of the Millimetron space observatory. The color scale shows the ratio of the expected flux from a galaxy similar to UGC~5101 at redshift $z$ (Fig.~\ref{fig-ugc5101-ions}) to the receiver sensitivity limit for an integration time of 1~h and spectral resolution $R = 10^3$ (Fig.~\ref{fig-ugc5101-ions}). 
}  
\label{fig-obspos-ugc5101-ions}
\end{figure}

Strong [CII] 158~$\mu$m emission may be observed in such galaxies up to $z\simlt 2$ (Fig.~\ref{fig-obspos-ugc5101-ions}). This line arises in warm neutral gas \citep{Wolfire1995,Kirsanova2020,Wolfire2022} and is a reliable tracer of star formation rate \citep{Kennicutt2012}. It should be noted that joint measurements of the underlying dust continuum allow obtaining constraints on the impact of X-ray radiation from the central SMBH \citep{Decarli2018,vsn2022muto,vsn2023fir}.

Figure~\ref{fig-obspos-ugc5101-ions} also shows that at redshifts $z\sim 0.5-2$ it becomes possible to observe other lines excited under similar physical conditions: [CI] 370~$\mu$m, [OI] 145~$\mu$m, as well as [NII] 122~$\mu$m, [NII] 205~$\mu$m и [OIII] 88~$\mu$m in warm ionized gas. Some studies of these lines have been carried out for lensed galaxies at $z\sim 1-3$ \citep{Zhang2018}. The fluxes of some of them will be detectable in galaxies similar to UGC 5101, with total IR luminosity $10^{12}\lsun$ using receivers of the Millimetron observatory. It can be seen that some are below the sensitivity limits by a factor of $\sim 10-100$ (the color scale shows the logarithm of the ratio of the expected galaxy flux at redshift $z$ to the receiver sensitivity limit for an integration time of 1 h and spectral resolution $R = 10^3$). In that regard, it is worth noting that the luminosities of some ultraluminous IR galaxies reach $\simlt 10^{13}\lsun$ \citep{Casey2014} and much brighter IR galaxies have been found at $z\sim 2.5-4.5$, with luminosities tens and hundreds of times higher \citep{Toba2016} up to $\sim 3.5\times 10^{14}\lsun$ \citep{Tsai2015}.

The observations of quasars at redshifts $z\simgt 6$ have shown that their host galaxies are rich in metals and dust, with IR and [CII] 158~$\mu$m line luminosities comparable to local ULIRGs \citep{Venemans2017a,Venemans2017b,Venemans2017c,Venemans2019,Decarli2018}. Intense star formation and massive black hole activity in these galaxies should produce strong [OIII] 88~$\mu$m and [NII] 122~$\mu$m lines, which at such high redshifts fall within the spectral range of the Millimetron receivers, whose sensitivity is sufficient to detect these lines even in lensed galaxies at $z\simgt 6$, including those having low metallicity \citep{Larchenkova2022}. It should be noted that growth of SMBHs and their influence on the ISM in galaxies at $z\sim 7-10$ can also be traced through continuum observations (see below), particularly bremsstrahlung emission in galaxies with primordial composition \citep{vs2019smbh} and dust emission in metal-enriched galaxies \citep{vsn2022smbh}.

Studying galaxies like UGC 5101 at high redshifts is challenging even with the ALMA interferometer. Figure~\ref{fig-ugc5101-ions} shows that the corresponding line fluxes are at the ALMA’s sensitivity limit. However, starting from $z\simgt 1$ hyperluminous IR galaxies begin to appear with luminosity $L_{IR}\simgt 10^{13}\lsun$ \citep{RowanRobinson2000,Toba2016}, which is an order of magnitude higher than UGC 5101. In recent years, hundreds of such galaxies have been discovered at $z\sim 1.1-3.3$ in the data from the WISE, Planck, Herschel, and South Pole Telescope (SPT) missions \citep{Assef2015,Tsai2015,Berman2022,Kamieneski2024}. These galaxies can be studied by the Millimetron observatory and ALMA array. Notably, at high redshifts, gravitational lensing becomes possible, and even rare objects such as hyperluminous IR galaxies can be strongly magnified. Among the detected sources, about 20 hyperluminous galaxies exceed $10^{14}\lsun$\citep{Berman2022,Kamieneski2024}. In particular, the luminosity of the lensed galaxy PJ0116-24 at $z\simeq 2.1$ reaches $\mu L_{IR} \sim 2.6\times 10^{14}\lsun$ for a magnification factor $\mu\simeq 17$ \citep{Liu2024}. At this redshift, bright lines such as [SIII] 33~$\mu$m, [SiII] 34~$\mu$m, [OIII] 52~$\mu$m и [NIII] 57~$\mu$m which are observed in local luminous IR galaxies such as NGC 6240 and UGC 5101, may be detectable with the Millimetron observatory. Detecting a greater number of lines will allow better understanding of ionization sources in the ISM of luminous IR galaxies \citep{Spinoglio1992}.

\section{MOLECULAR GAS}

The interstellar medium in luminous IR galaxies is subject to markedly more extreme influences than in most quiescent galaxies, including our own. In ultraluminous galaxies, there are favourable conditions for the formation of a wide variety of molecular species, as in Arp~220 \citep{Rangwala2011}, UGC~5101 \citep{CruzGonzalez2020}, and others \citep{Brauher2008,Liu2017}. Detection of a greater number of transitions in molecular lines will provide more information on the physical state of gas under the extreme conditions of the interstellar medium of ultraluminous IR galaxies.

\begin{figure}
\center
\includegraphics[width=170mm]{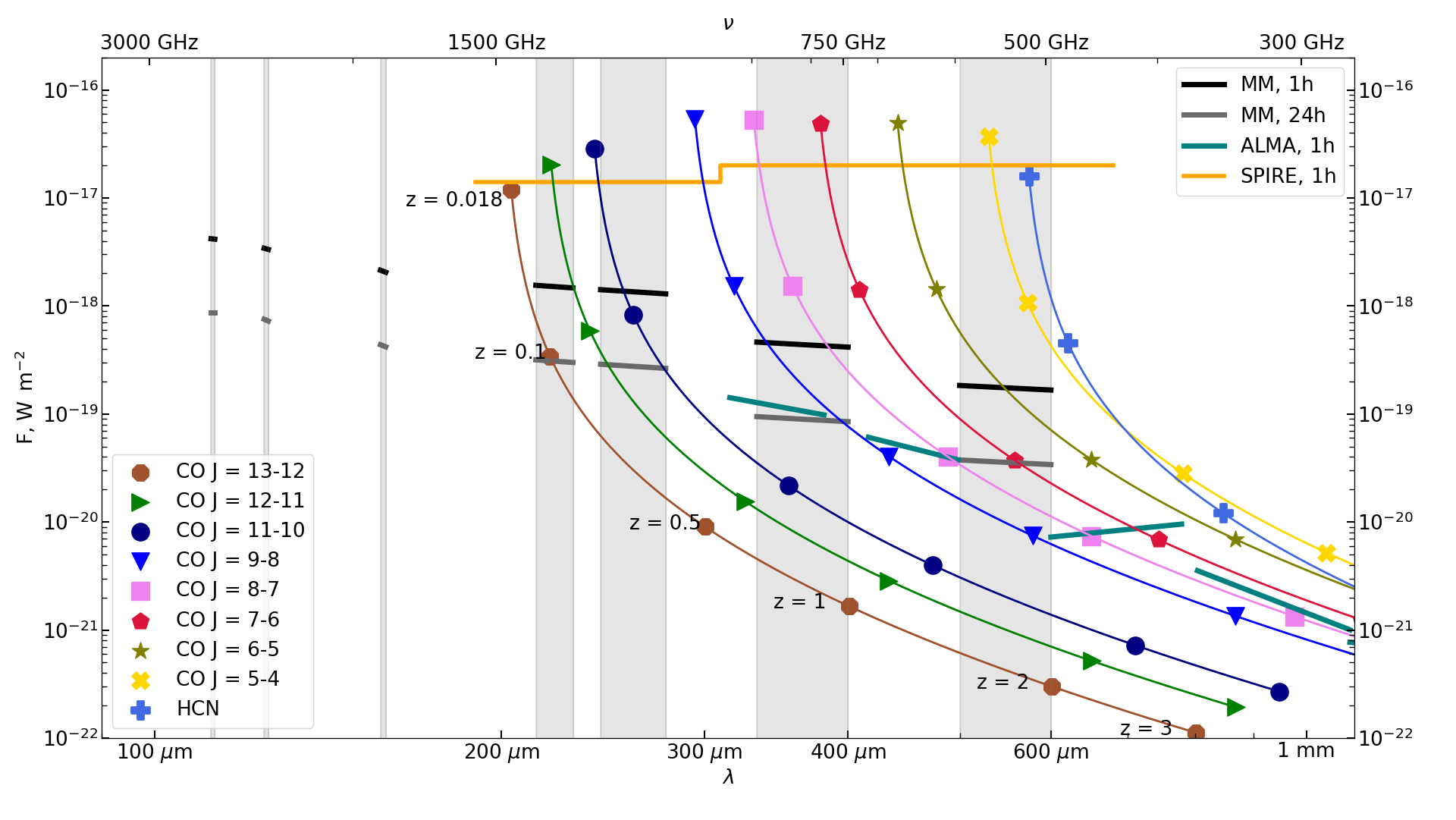}
\caption{ 
Expected fluxes in the CO and HCN molecular lines from a galaxy similar to Arp 220 at redshift $z$. The symbols from top to bottom correspond to the redshift of the galaxy Arp 220 $z=0.018$ and further $z=0.1$, 0.5, 1, 2 and 3 (for reference these values are indicated near the symbols along one of the curves). The fluxes are normalized to the total luminosity in the IR range $10^{12}\lsun$. Other notations are the same as in Fig.~\ref{fig-arp220-ions}.
}  
\label{fig-arp220-co}
\end{figure}

In molecular clouds of the Galaxy, the physical conditions favor the excitation of only the lower levels of CO molecules. Intense UV and X-ray radiation from starforming regions and the central SMBH stimulate to populate higher levels of CO molecule. Figure~\ref{fig-arp220-co} shows the expected line fluxes of CO from a galaxy similar to Arp 220. The transitions with $J>5$ can be observed in the ISM of luminous IR galaxies in the local Universe, and with $J>7$ -- at $z\simlt 0.5$. Simultaneous detection of several transitions will allow us to estimate density, temperature, and gas metallicity. Of particular importance is the possibility of observations of interstellar gas exposed to strong ionizing radiation fields and cosmic rays, especially transitions between highly excited levels. These conditions raise a question of the applicability of the ground CO transition intensity as a tracer of molecular hydrogen column density \citep{Bolatto2013,vkk2016} and the possible modification of the conversion factor in the extreme ISM of powerful starforming regions in luminous IR galaxies.

\subsubsection*{Water}

The abundance of water in the gas phase of molecular clouds in the Milky Way is relatively low, $X({\rm H_2O}) \simlt 10^{-9}$ \citep{Caselli2010}. However, water becomes one of the most widespread species behind shock fronts \citep{Bergin2003,GonzalezAlfonso2013} and in warm dense gas, where the stellar radiation causes ice to evaporate from dust grains \citep{Cernicharo2006}. Thus, water molecules can serve as an indicator of gas associated with regions of intense star formation and heated under the extreme conditions around active galactic nuclei. Due to their complex energy level structure, water molecules have numerous rotational lines in the far-IR. The water lines provide insight into the properties of the gas and the IR dust emission \citep{Weiss2010,GonzalezAlfonso2012,GonzalezAlfonso2014}. Highly excited water lines can be used to identify regions of strong IR absorption in galactic nuclei \citep{vanderWerf2011}. This allows distinguishing contributions from nuclear activity and star formation.

Ultraluminous IR galaxies turn out to be powerful emitters in water lines. Their total IR luminosities and water line luminosities are linearly correlated, and galaxies at high redshift have been found with extreme values exceeding $10^{13}\lsun$ and $10^8\lsun$, respectively \citep{Yang2013,Yang2016}.

Constraints on physical parameters can be improved by detecting related chemical components involved in water chemistry, in particular, OH lines and the molecular ions OH$^+$, H$_2$O$^+$, H$_3$O$^+$, observed in spectra of ultraluminous IR galaxies \citep[e.g.,][]{Rangwala2011,GonzalezAlfonso2013}, as well as other components such as HCN, HCO$^+$, NH, CH$^+$ within the ranges of the Millimetron observatory recivers. It should be noted that OH rotational transitions at 79, 119, and 163~$\mu$m are often seen in emission in galaxies with active nuclei, such as NGC 1068 \citep{Spinoglio2005}, whereas in luminous IR galaxies they are detected in absorption, as in Arp 220 \citep{Rangwala2011}. Apparently, the transition to higher IR luminosities is accompanied by an increase in optical depth, and molecular absorption lines provide another tool for probing ISM conditions (see discussion below). 

High transitions in water -- the "warm" and "hot" water transitions \citep[see Fig.3 in][]{Liu2017}, in particular ortho-${\rm H_2O}~3_{21}-3_{12}$ 1163 GHz, para-${\rm H_2O}~4_{22}-4_{13}$ 1208~GHz, ortho-${\rm H_2O}~6_{25}-5_{32}$ 1322 GHz, para-${\rm H_2O}~7_{44}-8_{17}$ 1345~GHz, para-${\rm H_2O}~4_{04}-3_{13}$ 2392~GHz, which fall into the M3, M4, and M6 bands of the high-resolution spectrometer of the Millimetron observatory \citep{Kirsanova2025} -- are excited in dense, hot molecular gas exposed to strong UV/X-ray radiation and cosmic rays. Such conditions exist in the central regions of galaxies with high SFRs (e.g., M82), near active galactic nuclei (NGC 1068, Mrk 231, etc.), and in ultraluminous IR galaxies (Arp 220 and others). Intense IR emission in the para-${\rm H_2O}~4_{22}-4_{13}$ line from galaxies Mrk 231 and Arp 220 should correspond to regions with significant masses of hot ($T_K$ and $T_{dust} \sim 100-200$~K) and dense $n_{\rm H} \simgt 10^6$cm$^{-3}$) gas with high column density $N_{\rm H}\simgt 10^{24}$cm$^{-2}$. The size of the region can be estimated at $\sim 60-100$~pc \citep{Downes2007}. In such hot gas, the levels are populated up to energies of 1000~K due to effective collisional excitation combined with strong IR radiation and numerous mid-IR transitions in water molecules. The Herschel/HIFI survey has revealed not only emission, but also absorption data in a large number of water transitions in ultraluminous IR galaxies \citep{Liu2017}.

\begin{figure}
\center
\includegraphics[width=170mm]{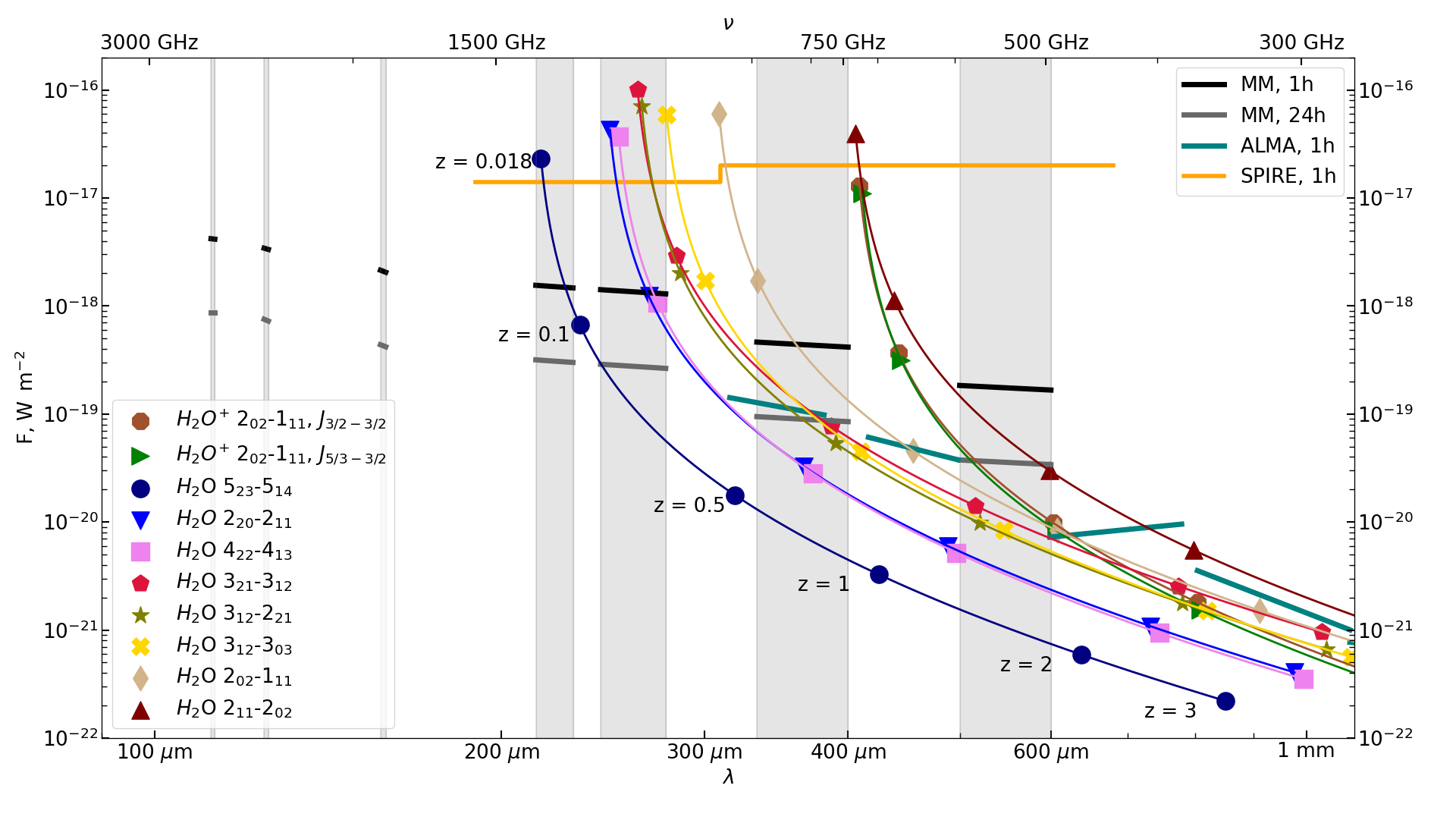}
\caption{ 
Expected fluxes in the H$_2$O from a galaxy similar to Arp~220, molecular lines from a galaxy similar to Arp~220 at redshift $z$. The symbols from top to bottom correspond to the redshift of the galaxy Arp~220 $z=0.018$ and further $z=0.1$, 0.5, 1, 2 and 3 (for reference these values are indicated near the symbols along one of the curves). The fluxes are scaled to the total IR luminosity $10^{12}\lsun$. Other notations are the same as in Fig.~\ref{fig-arp220-ions}.
}  
\label{fig-arp220-h2o}
\end{figure}

Figure~\ref{fig-arp220-h2o} shows the expected fluxes in the H$_2$O lines from a galaxy similar to Arp 220. Thus, in the scientific program of the Millimetron observatory, it appears feasible to study the excitation conditions of "warm" and "hot" water transitions in luminous IR galaxies up to $z\simlt 0.5$. It is important that, fluxes in H$_2$O$^+$ ion lines are sufficient for detection in the same range of redshifts. This molecular ion participates in the water formation chemistry and is a good tracer of cosmic-ray and X-ray ionization rates. Joint analysis with line ratios of atomic fine-structure lines of metal ions (see previous section) will help constrain the physical properties of the ISM in ultraluminous IR galaxies \citep[e.g.,][]{Meijerink2005,Meijerink2007} and determine the possible thermal structure of the medium through excitation of various water transitions.

\subsubsection*{The Effect of Mixing}

It has long been known that population synthesis and photoionization models for ultraluminous IR galaxies suggest that short starburst episodes, only a few million years old, account well for their spectral properties \citep{Satyapal2008,Colbert1999,Fischer1996}. The gas surface density and the supernova rate in star-forming regions of IR galaxies exceed those in the Milky Way by 3–4 orders of magnitude \citep{Kennicutt2012}.

It can be said that chemical inhomogeneities persist in the medium if the metal mixing timescale exceeds the dynamical timescale. The latter can have an upper limit, for example, for an active region in Arp 220 of size $R_{\rm IR}\sim 1$~kpc \citep{Emerson1984} and a characteristic merging velocity $v_m\geq 100$~km~s$^{-1}$ yielding dynamical time $t_{dyn}\leq 10$ Myr. Mixing efficiency is governed by turbulent diffusion \citep{Avillez2002} and the rate of strong shock crossing an ISM gas element \citep{Pan2010}: $\nu_s(v) \sim 5\times 10^{-5}\nu_{SN}v^{-1}$~yr$^{-1}$, where $v$ is shock velocity in km~s$^{-1}$ and $\nu_{SN}$ is the supernova rate. Thus, for ultraluminous IR galaxies, the interval between subsequent shocks is about $\nu_s^{-1}\sim 1$~Myr for shock velocities of $\sim 100$~km~s$^{-1}$. However, one should bear in mind that supernovae explode in clusters. Under such conditions, the timescale for the merging of regions around individual stellar clusters may be on the order of tens of millions of years \citep{vns2015}. This timescale characterizes the loss of chemical inhomogeneity in the ISM -- metal mixing \citep{Avillez2002}. The spatial scale of such inhomogeneities is apparently of the order of stellar cluster sizes, \citep{Krumholz2019}. 

Thus, different ISM regions in ultraluminous IR galaxies can remain chemically distinct, and elemental abundances can vary spatially. This has a significant influence on the thermal evolution of interstellar gas and the ratios of emission line intensities of metals \citep{vs2016,vs2017} and molecular composition \citep{vv2025}. For example, hydroxyl molecules are key species in the water formation chain. It is worth noting, however, that the spatial resolution of the Millimetron observatory will be much worse than the characteristic scale of such inhomogeneities, but local variations in the line emission intensity may still yield a cumulative effect. 

\begin{figure}
\center
\includegraphics[width=170mm]{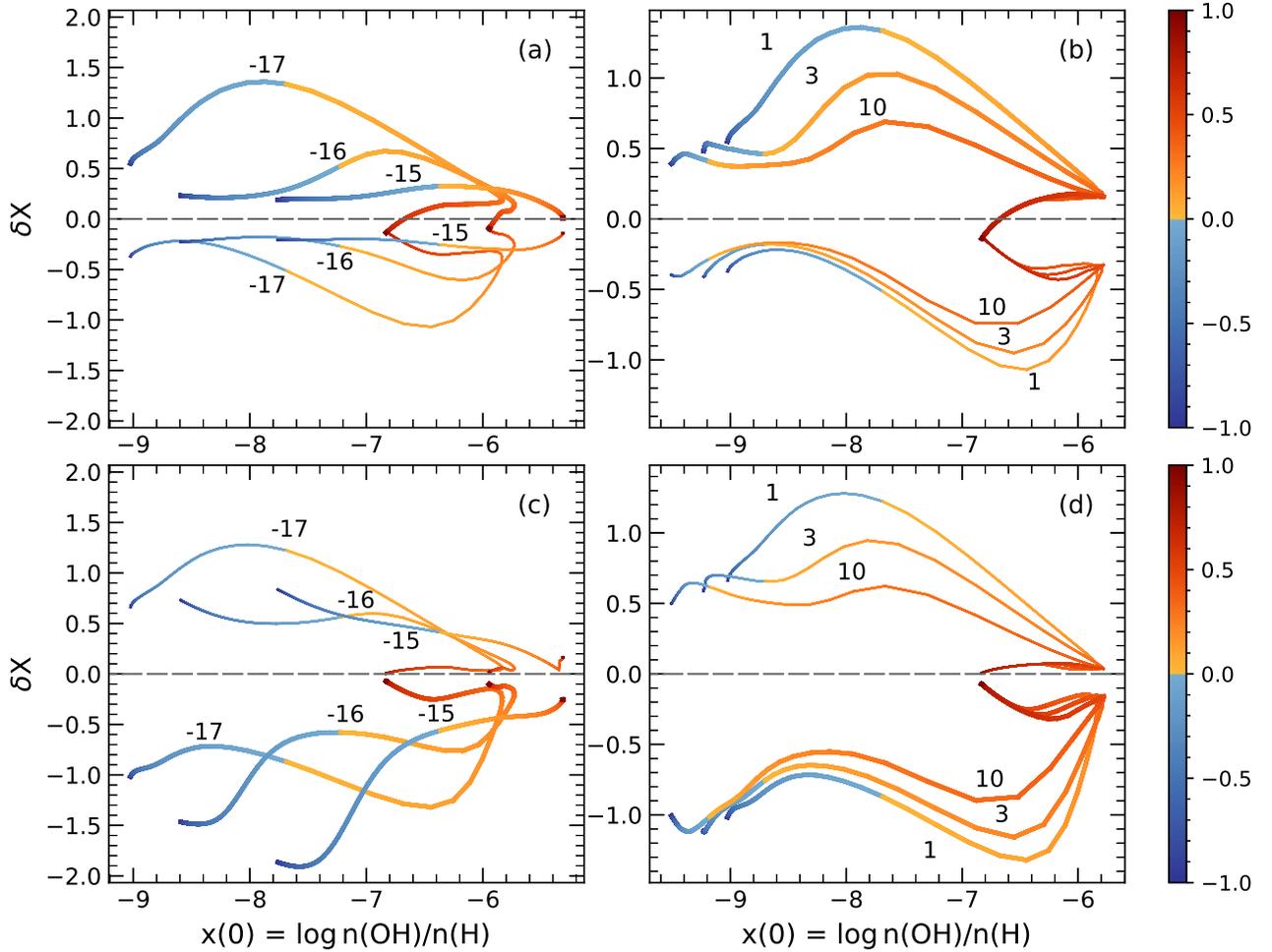}
\caption{ 
Relative change of the OH molecule fraction for variation of oxygen abundance $\delta{\rm[O/H]} \pm 0.3$~dex (top row) and carbon $\delta{\rm[C/H]} \pm 0.3$~dex (bottom row) in gas with ISM elemental composition for different values of cosmic ray ionization rate (left column) and UV radiation flux (right column). Thick lines correspond to an increase in the abundance by 0.3~dex, thin lines to a decrease by 0.3~dex. Panels (a) and (c) show the dependence for ionization rate values from $\zeta_{\rm H_2} = 10^{-17}$ to $10^{-15}$~s$^{-1}$ (labels correspond to the logarithm of the value) at fixed UV flux. Panels (b) and (d) show the dependence for UV flux values from $G_0$ to $10\, G_0$, labels correspond to flux values in Habing units $G_0$ \citep{Habing1968} for fixed ionization rate $\zeta_{\rm H_2} = 10^{-17}$~s$^{-1}$. The gas density is $n_{\rm H} =10^4$ cm$^{-3}$. The color scale indicates the logarithm of the extinction ${\rm log} A_V$.
}  
\label{fig-var-oh}
\end{figure}

Figure~\ref{fig-var-oh} shows the relative change of the OH abundance for a variation of oxygen and carbon abundance $\delta X = x(\delta{\rm[X/H]})-x(0)$ from its value in the typical gas composition for the ISM, for different cosmic ray ionization rates and UV fluxes. The horizontal line $\delta X = x(\delta{\rm X})-x(0) = 0$ corresponds to zero deviation from the OH fraction in gas with the typical composition for the Galactic ISM \citep{Sembach2000}. It is evident that the variations in oxygen and carbon abundances lead to the difference in the OH fractions by a factor of several times. These deviations are especially pronounced for low extinction $A_V$ (the color scale shows the logarithm of extinction). Increasing ionization rate and UV flux the differences decrease, but remain significant. Under such conditions the OH fraction rises reaching ${\rm [OH/H]} \sim -6$, with maximum values corresponding to high extinction. This may manifest as strong OH line absorption, and incomplete mixing enhances it further.

\section{DUST}

Dust is produced in the ejecta expelled during Type II supernova explosions in the free-expansion stage \citep{Todini2001,Nozawa2006}. On the other hand, as a supernova remnant expands, the swept-up interstellar dust is destroyed by strong shock waves \citep{Jones1996,Bocchio2014,Micelotta2016} through processes of thermal and kinetic sputtering in hot gas with $T\simgt 10^6$К \citep{Barlow1978,Draine1979a,Draine1979b} and shattering in collisions between dust grains in dense regions \citep{Borkowski1995,Jones1996,Bocchio2016}. It should be noted that large grains $a\simgt 0.1$~$\mu$m can penetrate far beyond the shock front into hotter ejecta gas, remain there for tens of thousands of years, and be efficiently destroyed during this period \citep{Slavin2020,vs2024}.

In luminous IR galaxies the supernova rate reaches from several to dozens per year \citep{Kennicutt2012,Varenius2019}. At such high rates, in the central region of a galaxy one can expect individual remnants to overlap (merge with one another) before the end of the Sedov–Taylor phase, which may result in both efficient dust destruction in the surrounding interstellar medium and the onset of a large-scale wind with high heating efficiency of the interstellar medium \citep{Melioli2004,vns2015,vsn2017}, and consequently, the outflow of dust from this region.

A high rate of supernova explosions support strong turbulence in the medium and the formation of numerous dense fragments during the disruption of bubble shells \citep{vns2015,Li2015}. For cloud filling factor $f_v\simeq 10^{-2}-10^{-3}$, most of the dust is apparently confined in molecular clouds. Expanding in a cloudy medium, the supernova shell interacts with gas of varying density: the shock wave penetrates more rapidly into low-density regions and decelerates in dense clouds \citep{Korolev2015,Slavin2017,Wang2018}. In this case, the bulk of the dust contained in molecular clouds should not be destroyed by shock waves \citep{Dedikov2024u,Dedikov2024b} and should emit at wavelengths of 100--500~$\mu$m \citep[e.g.,][]{Drozdov2025}.

To maintain high IR luminosity, dust should be efficiently heated by emission from the stellar population and the active nucleus. Evidently, an old stellar population aged several hundred million years does not contribute significantly to the heating, since it does not contain stars with a mass more than $8\msun$. The observed star formation rates in luminous IR galaxies most likely do not provide the required number of ionizing photons either \citep{sv2017}. In this case, it can be assumed that a significant fraction of the young population is embedded in the central region with high extinction and is therefore hidden from observations. This is supported by the observations of emission in the HCN and HCO$^+$ lines in the circum-nuclear region, as well as in highly excited rotational levels of CO (from $J=4-3$ to $13-12$) and other molecules \citep{Rangwala2011}. Favorable conditions for exciting such states arise in dense gas exposed to strong ionizing radiation and cosmic rays from the hidden young population. There are indications that a significant fraction of the energy release originates not from the observed stellar population, but from hidden sources -- stars, supernovae, and the active nucleus \citep{Varenius2019}.

The estimates of the number of supernovae in the IR-dominant region may be underestimated both due to strong absorption in the IR \citep[optical depth can reach $\sim 5$ at 100~$\mu$m,][]{Rangwala2011} and due to the strong magnetic field (up to 10 mG) in this region \citep{Batejat2011}, which shortens the cooling time of cosmic-ray electrons to $\sim 3\times 10^{13}\gamma^{-1}$~s, corresponding to only about a hundred years for a gamma factor of $\gamma\sim 3\times 10^3$ \citep{sv2017}.

Using measurements of the spectral flux of dust IR emission, the dust mass \citep{Hildebrand1983} and equilibrium dust temperature can be derived. Both values depend on the precise value of the spectral index $\beta$ of the dust absorption coefficient. For characteristic dust masses of $\sim 10^8\msun$ and total IR luminosity of $10^{12}\lsun$ at standard $\beta=2$, the dust temperature is estimated at $T_d\sim 36$~K, while for $\beta=1.83$ it is $T_d\sim 59$~K \citep{Rangwala2011}. Thus, observations of dust IR emission in luminous IR galaxies provide constraints on dust properties, heating sources, and physical conditions in the interstellar medium. The current  measurements of the dust continuum in luminous IR galaxies \citep{Brauher2008,Rangwala2011} suffer from insufficient angular resolution, preventing separate study of emission from the central IR-dominant region.

\begin{figure}
\center
\includegraphics[width=170mm]{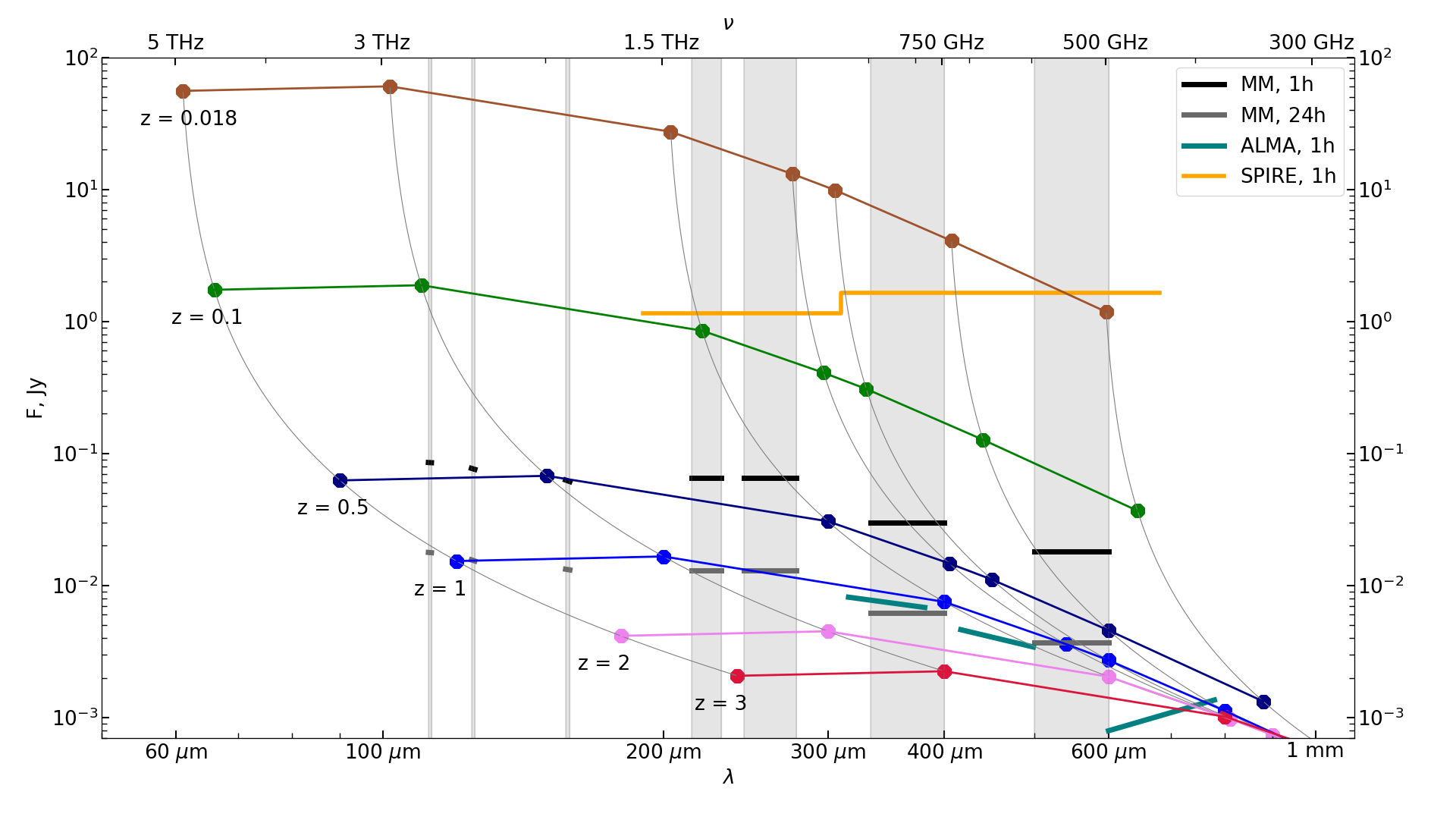}
\caption{ 
Expected fluxes in the dust continuum from a galaxy similar to Arp 220 at redshift $z$. { The symbols from top to bottom correspond to the redshift of the galaxy Arp 220 $z=0.018$ and further $z=0.1$, 0.5, 1, 2 and 3 (for reference these values are indicated near the symbols along one of the gray curves)}. The fluxes for $z=0.018$ correspond to those observed from the galaxy Arp~220 and normalized to the total IR luminosity $10^{12}\lsun$ \citep{Brauher2008,Rangwala2011}. Gray vertical bands show the wavelength ranges of the high-resolution spectrometer of the Millimetron observatory. Thick black lines correspond to the $5\sigma$ receiver sensitivity for an integration time of 1~h and spectral resolution $R = 300$. Thick gray lines correspond to 24~h of integration. Thick green lines show the sensitivity for the ALMA interferometric array observations in the 50-antenna mode at $R=300$ and 1~h integration, with significance of $5\sigma$. The thick yellow line corresponds to the Herschel/SPIRE sensitivity at 1~h integration with resolution $R=10^3$. 
}  
\label{fig-arp220-dust}
\end{figure}

The dust continuum of IR galaxies like Arp 220 can be studied using the Millimetron observatory up to redshifts of $\sim 0.3$ with 1~h integration time, and nearly to $z \sim 1$ in the case of 24~h integration time (Fig.~\ref{fig-arp220-dust}). It can be seen that the spectral peak almost always lies within the range of the Millimetron spectrometer. Measurements of the long wavelength part of the spectrum will provide constraints on dust temperature, thereby enabling studies of dust properties and heating sources.

\subsubsection*{Absorption Lines}

Observations of molecular absorption lines against the dust continuum indicate powerful outflows with velocities up to several thousand km/s from active galaxies \citep{Veilleux2013}. The origin of such molecular outflows and the survival of molecules under their extreme conditions remain unresolved problems. The absorption in OH lines at 119, 79, 84, and 65~$\mu$m in Mrk 231 \citep{GonzalezAlfonso2014}, in water lines, and in other molecules in NGC~4418 and Arp~220 \citep{GonzalezAlfonso2012}, is significant. Apparently, this is sufficient to study molecular outflows from galaxies using the high-resolution spectrometer on board the Millimetron observatory at redshifts $z\sim 1-2$ in water and hydroxyl lines. The discovery of several dozen hyperluminous lensed IR galaxies \citep{Liu2024} with luminosities $\sim 10-100$ times higher than Arp 220 will allow studying outflows from these objects. It is known that unambiguous signatures of outflows include P Cyg-type profiles or absorption in the wings of OH spectral lines, observed in several nearby ultraluminous IR galaxies \citep{GonzalezAlfonso2017}. The absorption and emission peaks in the profiles of these galaxies are separated by nearly 1000 km/s. Therefore, a spectral resolution $R=300$ is insufficient for outflow velocities of $\sim 1-2$ thousand km/s. For more reliable conclusions, measurements with $R\simgt 10^3$ are desirable. From the shapes of these lines, one can estimate the energy input powering these outflows, thus obtaining independent constraints on the star formation rate and the SMBH mass in IR galaxies.

\subsubsection*{Magnetic Fields}

Magnetic fields are an important component of the interstellar medium, influencing star formation, the transport of gas and dust, the structure of the medium, and the accretion rate onto SMBHs \citep{Beck2013}. Dust grains can align under the influence of magnetic fields, leading to polarization of the IR and submillimeter radiation. Sensitivity and angular resolution are important in studies of dust polarization; therefore, investigations of the dust continuum in ultraluminous IR galaxies with the Millimetron observatory can significantly improve our understanding of magnetic-field structures in such objects. 

The measurements of dust polarization in Arp 220 with the SCUBA millimeter array provided only the upper limit of about 1.54\% at 850~$\mu$m, averaged over a beam size of 15'' \citep{Seiffert2007}. This angular scale is much larger than the separation between regions surrounding the active nuclei in the galaxy. If the polarization differs in these regions, it may be diluted to a low level. 

Shifting to shorter wavelengths leads to higher optical depth and improved spatial resolution. The optical depth at 850~$\mu$m has been estimated at $\sim 1-3$ \citep{Sakamoto2008}, while at 100~$\mu$m it reaches $\sim 5$ \citep{Rangwala2011}. For the Millimetron 10-m mirror, the angular resolution will be 3'' at a wavelength of 100~$\mu$m, which allows us to separate emission from each nucleus. Measurements of dust polarization at several frequencies in the 100--500~$\mu$m range with the Millimetron observatory will provide constraints on dust models and magnetic-field strengths in the ISM of nearby IR galaxies. Notably, the recent SMA millimeter-array measurements have revealed dust polarization in Arp 220 at a level of 2.7\% in the western nucleus region \citep{Clements2025}.

\section{CONCLUSIONS}

The conditions in the interstellar medium of ultraluminous IR galaxies differ substantially from those in galaxies with quiescent star formation, such as our own, due to the presence of intense fluxes of X-ray radiation, cosmic rays, and high dust fraction. In this respect, they resemble galaxies from the epoch before hydrogen reionization -- a time of extremely rapid growth of SMBHs. Studying these local ''laboratories'' will help constrain the properties of the interstellar medium typical for the emergence of conditions that give rise to intense star formation and efficient accretion onto SMBHs.

The spectral capabilities of the Millimetron observatory will make it possible to address a number of questions regarding the emergence of galaxies with extremely high infrared luminosity, the properties of gas and dust in the star-forming regions of these objects, and to gain a better understanding of the following:

\begin{itemize}

 \item the possible evolutionary connection between local ultraluminous IR galaxies and dusty galaxies at high redshifts;

 \item the reasons for extremely high IR emission and the sources of dust heating;

 \item the kinetics of water molecules, carbon monoxide, and associated molecular ions in the dense clouds of IR galaxies;

 \item the variations in the estimates of molecular hydrogen mass and the values of the conversion factor under conditions of high SFRs;

 \item the sources of dust in ultraluminous infrared galaxies, as well as the processes of dust production and destruction in environments with highly efficient shock-wave heating;

 \item the nature of molecular outflow emergence in galaxies with high SFRs;

 \item the effective influence of magnetic fields on the dynamics and emission properties of dust particles.

\end{itemize}

The progress in these areas will lead to significant refinements and possibly a revision of our understanding of the processes in the interstellar medium that give rise to conditions for extremely high SFRs and the growth of SMBHs.

\section*{ACKNOWLEDGEMENTS}
The authors thank the anonymous reviewer for valuable comments and S.V. Pilipenko for fruitful discussions.

\section*{FUNDING}
This work was supported by ongoing institutional funding. No additional grants to carry out or direct this particular research were obtained.

\section*{CONFLICT OF INTEREST}

The authors of this work declare that they have no conflicts of interest.

\bibliographystyle{aspb1}
\bibliography{ulirgs-bib}

\end{document}